\documentstyle[12pt,aasms4] {article}
\newcommand{\ts}{\thinspace}

\begin{document}

\title{Near-Infrared Spectroscopy and a Search for CO Emission in Three 
Extremely Luminous {\it IRAS} Sources; IRAS F09105+4108, IRAS F15307+3252, and
PG 1634+706}

\author{A. S. Evans\footnote{Present Address: Dept. of Astronomy 105-24,
California
Institute of Technology, Pasadena, CA 91125} 
$^{,}$\footnote{Guest Observer at the James Clerk Maxwell
Telescope, which is operated by the Royal Observatory Edinburgh on behalf
of the United Kingdom Science and Engineering Research Council (SERC),
the Netherlands Organization for the Advancement of Pure Research (ZWO), and
the Canadian National Research Council (NRC).} and D. B. Sanders}
\affil{Institute for Astronomy, 2680 Woodlawn Drive, Honolulu, HI 96822;
ase@astro.caltech.edu, sanders@galileo.ifa.hawaii.edu}

\author{R. M. Cutri}
\affil{IPAC, MS 100-22, California Institute of Technology,
Pasadena, CA 91125;
roc@ipac.caltech.edu}

\author{S. J. E. Radford}
\affil{National Radio Astronomy Observatory\footnote{
The NRAO is a facility of the National Science Foundation operated 
under cooperative
agreement by Associated Universities, Inc.}, Tucson, AZ 85721;
sradford@nrao.edu}

\author{J. A. Surace}
\affil{Institute for Astronomy, 2680 Woodlawn Drive, Honolulu, HI 96822;
jason@galileo.ifa.hawaii.edu}

\author{P. M. Solomon}
\affil{Astronomy Program, State University of New York, Stony Brook, NY 11794;
psolomon@sbast7.ess.sunysb.edu}

\author{D. Downes}
\affil{Institut de Radio Astronomie Millim\'{e}trique, 38406 St. Martin
d'H\`{e}res, France;
downes@iram.fr}

\author{C. Kramer}
\affil{Physikalishes Institut, Universitaet zu Koeln, Zuelpicher
Strasses 77, D-50937 Koeln, Germany;
kramer@ph1.uni-koeln.de}

\clearpage

\begin{abstract}

Rest-frame 0.48--1.1 $\mu$m emission line strengths and molecular gas mass
(H$_2$) upper limits for three luminous infrared sources -- the
hyperluminous infrared galaxies\footnote{HyLIGs: $L_{\rm ir} \geq 10^{13}
L_\odot$ where $L_{\rm ir} \equiv L(8-1000\mu m)$ and assuming $H_0 =
75{\ts}$km s$^{-1}$ Mpc$^{-1}$ and $q_0 = 0$ (e.g. Sanders \& Mirabel
1996). Additional infrared galaxy definitions used throughout this paper
-- Luminous Infrared Galaxies (LIGs): galaxies with $L_{\rm ir} =
10^{11-11.99} L_\odot$, Ultraluminous Infrared Galaxies (ULIGs):  galaxies
with $L_{\rm ir} \geq 10^{12} L_\odot$.  Throughout this paper, we adopt
$H_0 = 100h{\ts}$km s$^{-1}$ Mpc$^{-1}$ and $q_0 = 0.5$.}
IRAS{\ts}F09105+4108\footnote{The source prefix `F' will be used
throughout this paper as a shorthand to refer to sources listed in the
IRAS Faint Source Catalog (FSC: Moshir et al. 1992).}
 ($z=0.4417$), IRAS{\ts}F15307+3252 ($z=0.926$), and the
optically-selected QSO PG{\ts}1634+706 ($z=1.338$) -- are presented.
Diagnostic emission-line ratios ([O III]{\ts}$\lambda$5007/H$\beta$, [S
II]{\ts}$\lambda \lambda$6716, 6731/H$\alpha$, [N
II]{\ts}$\lambda$6583/H$\alpha$, and [S III]{\ts}$\lambda
\lambda$9069+9532/H$\alpha$) indicate a Seyfert 2-like spectrum for both
infrared galaxies, consistent with previously published work.  The upper
limits of molecular gas mass for all three sources are $M({\rm H}_2) < 1-3
\times 10^{10} h^{-2}\ M_\odot$ ($q_0=0.5$, $H_0 = 100h$ km s$^{-1}$
Mpc$^{-1}$), less than that of the most gas-rich infrared galaxies in the
local Universe.  All three sources have $L_{\rm ir}/L'_{\rm CO} \sim
1300-2000$, and thus are the extragalactic sources with the most extreme
$L_{\rm ir}/L'_{\rm CO}$ values measured to date. Given the relatively
warm far-infrared colors for all three objects, much of their infrared
luminosity may emanate from a relatively modest amount of warm dust (e.g.
$M_{\rm d} \sim 10^{5-7}\ M_\odot$, $T_{\rm d} = 200-100${\ts}K) near the
AGN.  For F09105+4108 and F15307+3252, the implied circumnuclear covering
factor of this dust is $\sim 90$\%, while for PG{\ts}1634+706 the covering
factor is only $\sim 35$\%.

\end{abstract}


\section{Introduction}

Studies of ultraluminous infrared galaxies (ULIGs) detected by {\it IRAS}
have focussed on revealing the nature of the host galaxies and the source
powering the enormous infrared luminosities.  Three forms of spectroscopic
evidence for both stellar and nonstellar energy sources exists:  (1)
Studies of ULIGs comparing the fluxes of hydrogen recombination lines with
optical and near-infrared forbidden lines have shown that the dominant
ionization process appears to differ from galaxy to galaxy.  Emission-line
diagnostics have been used to classify these galaxies as H II region-like
galaxies (i.e., emission lines induced by hot OB stars), Low Ionization
Nuclear Emission-line Regions galaxies (LINERs), or Seyferts (e.g.,
Sanders et al. 1988b; Kim et al. 1995; Kim 1995; Veilleux et al. 1995).
(2) Spectral polarimetry of ULIGs often reveal highly polarized continuum
and broad emission-line spectrum, indicating the presence of a buried AGN
(e.g., Kay \& Miller 1989; Hines 1991; Hines 1994; Hines \& Wills 1993;
Jannuzi et al. 1994).  (3) Millimeter-wave (CO) spectroscopy of ULIGs has
shown them to be rich in molecular gas, indicating the presence of an
enormous amount of material for fueling an active galactic nucleus (AGN)
and star formation (e.g., Young et al. 1984; Sanders and Mirabel 1985;
Sanders et al. 1987; Scoville et al. 1986; Sanders, Scoville, Soifer 1991;
Sanders 1991).  In addition, morphological studies of ULIGs have revealed
that they commonly have tidal distortions and nearby companions indicative
of the merger of two or more galaxies (e.g. Sanders et al.  1988a,b).
Sanders et al. 1988a,b have suggested that ULIGs may be the precursor to
optically-selected quasars.

The recent discovery of the high-redshift HyLIG IRAS{\ts}10214+4724
($z=2.286$: Rowan-Robinson et al. 1991) and the subsequent detection of
molecular gas in the source (Brown \& Vanden Bout 1991, 1992; Solomon,
Downes, \& Radford 1992a), as well as evidence that it harbors a buried
AGN (e.g. Elston et al. 1994; Jannuzi et al. 1994), has sparked interest
in searching for and studying more luminous IRAS galaxies at early epochs
of the universe. To date, there are two other published findings of $z >
0.3$ galaxies discovered by {\it IRAS} -- F09105+4108 ($z=0.4417$:
Kleinmann \& Keel 1987; Kleinmann et al. 1988) and F15307+3252 ($z=0.926$:
Cutri et al. 1994), and surveys are underway to search for additional
candidates.  In addition, a significant fraction of moderate-to-high
redshift QSOs were detected by {\it IRAS}, making observations that test
the infrared galaxy-QSO evolutionary hypothesis at higher redshift (and
higher luminosities) feasible.  To date, 2 local QSOs (Mrk{\ts}1014:
Sanders, Scoville, \& Soifer 1988a; I{\ts}Zw{\ts}1: Barvainis, Alloin, \&
Antonucci 1989), one moderate-redshift quasar (3C{\ts}48 at $z=0.3695$:
Scoville et al. 1993; Wink et al. 1997), and 2 higher redshift QSOs (the
gravitationally lensed H{\ts}1413+117 at $z=2.556$: Barvainis et al. 1994;
BR{\ts}1202-07 at $z=4.695$: Ohta et al. 1996; Omont et al. 1996) have
been unambiguously detected in CO.

In this paper, we present CO observations and near-infrared spectroscopy
of the high redshift HyLIGs F09105+4108, F15307+3252, and PG{\ts}1634+706
(IRAS{\ts}16347+7037: e.g. Sanders et al. 1989), the latter of which is
the most infrared luminous, optically-selected QSO detected by {\it IRAS}
in the PG sample.  The purpose of these observations is to extend the
determination of molecular gas properties and dominant ionization
mechanisms to distant sources with extreme $L_{\rm ir}$, as well as to
establish whether their properties are consistent with the trends in such
properties observed for local ULIGs and QSOs.

This paper is divided into 5 sections.  The near-infrared and millimeter
observing procedures are discussed in \S 2. The data reduction methods and
line-flux calculations are summarized in \S 3.  Section 4 begins with a
brief discussion of the optical and infrared morphologies of the three
sources, then the dominant ionization mechanism in the two IRAS galaxies,
the interpretation of the $L_{\rm ir}$ and $L'_{\rm CO}$ of the 3 sources,
and their spectral energy distributions are examined.  Section 5
summarizes the results.  Throughout this paper, we adopt $H_0 = 100h$ km
s$^{-1}$ Mpc$^{-1}$ and $q_0 = 0.5$.

\section{Observations}

Tables 1--4 summarize basic information and a journal of observations for
the 3 sources discussed in this paper.  Observations with each of the
telescopes used are discussed separately below.  A more in depth
discussion of the observing technique and the difficulties inherent in
millimeter-wave and submillimeter-wave observations of faint sources can
be found in Evans et al. (1996).

\subsection{Imaging}

All of the imaging data were obtained at the UH 2.2m Telescope on Mauna
Kea. The near-infrared imaging data were taken with the UH QUick Infrared
Camera (QUIRC: Hodapp et al. 1996), which consists of a 1024x1024 pixel
HgCdTe Astronomical Wide Area Infrared Imaging (HAWAII) array.  The R-band
image of F15307+3252 and the I-band image of PG{\ts}1634+706 were taken
with the UH Tektronix 2048x2048 CCD camera (Wainscoat 1996).  Finally, the
I-band image PG{\ts}1634+706 was obtained with the Orbit Semiconductor
2048x2048 CCD camera (Wainscoat 1996), reimaged to f/10.  With the
exception of the R-band image of F15307+3252 which consisted of one long
exposure, all observations were done by taken a series of dithered
exposures.

\subsection{Near-Infrared Spectroscopy}

\subsubsection{UH 2.2m Telescope}

All spectroscopic observations on the UH{\ts}2.2m telescope were made with
the K-band spectrograph (KSPEC: Hodapp et al. 1994).  KSPEC is a
cross-dispersed echelle configured to cover the wavelength range 1.1--2.5
$\mu$m in 3 orders ({\it J, H,} and {\it K}) on a 256$\times$256 NICMOS-3
HgCdTe detector array.  The  wavelength resolutions (at 2.2 $\mu$m) for
each of the observations are listed in Table 3.  Additionally, the
wavelength range 0.7--1.0{\ts}$\mu$m is also dispersed in several orders
on the array. However, the order crowding at 0.7--1.0{\ts}$\mu$m is such
that only spectra of point-like sources can effectively be extracted, and
this wavelength range is also compromised due to the fact that KSPEC is
only in focus at $\lambda >${\ts}0.85{\ts}$\mu$m.  In addition to its
spectral capabilities, KSPEC also provides simultaneous imaging of
approximately one square arcminute of sky around the slit on a second
NICMOS-3 array.  For the 1994 December observing period, the new UH
tip-tilt system (Jim et al. 1997) was also implemented, notably improving
the tracking, offsetting, and seeing of the observations and data.

Each observing night began with a series of flat field exposures in the
closed dome with incandescent lights turned on, then off.  Once the
telescope was guiding with the source in the slit, a 180-second exposure
was taken, followed by a 180-second exposure 10 arcseconds off-source.
Each spectrum exposure was accompanied by a 140-second image exposure. For
the March observations, the pattern for observing was
source-sky-source-sky, etc., but the pattern was changed to
source-sky-sky-source-source-sky-sky-source, etc. for the December
observations to minimize spurious features created by the changing sky
conditions.  Observations of standard A{\ts}(V) stars near each source
were also made to flux calibrate the source spectrum and to remove
telluric lines, and observations of an argon lamp were made to wavelength
calibrate the KSPEC data.

\subsubsection{United Kingdom Infrared Telescope (UKIRT)}

Observations of F15307+3252 were made with the upgraded Cooled Grating
Spectrometer (CGS4) on the 3.8m UKIRT in shared-risk time.  CGS4 is a 1-5
$\mu$m 2D grating spectrometer with a 256$\times$256 InSb array.  Because
the H$\alpha$+[N II]{\ts}$\lambda$6583 and [S II]{\ts}$\lambda
\lambda$6716,6731 (hereafter [S II]{\ts}$\lambda$6724) emission from the
source is redshifted to J-band, the 75 line mm$^{-1}$ grating in second
order was used.  Sampling was done by moving the detector over 2 pixels in
6 steps in the wavelength direction. For the observations, a 1 pixel-wide
(1.2 arcsecond) slit was used.

Spectral data was taken in 120-second exposures, with 20 seconds of each
exposure made at each detector position during the 2-pixel sampling.  The
slit was then moved 15 arcseconds in the spatial axis direction of the
array for the next 120 second exposure. Thus, the source appears on the
array as a positive and negative spectrum after the first frame is
subtracted from the second.  The source-sky observation sequence was the
same as that used in the 1994 December KSPEC observations. Observations of
the standard A star HD{\ts}162208 were made before and after the
observations of F15307+3252.  A krypton lamp was used as the wavelength
calibration source.

We encountered a problem with CGS4; the source flux of the positive
spectrum was more than of the negative spectrum flux, implying that the
source was only partially in the slit in the second position.
Alternatively, the flux of the positive A star spectrum was less than the
negative spectrum flux. The slit also appeared to be tilted relative to
the array as evident by the position of the OH sky lines.  Because of
this, only half the data had enough flux to be usable, and no photometry
was possible.

\subsection{Submillimeter Spectroscopy} \subsubsection{James Clerk Maxwell
Telescope (JCMT)}

Observations with the 15m JCMT Telescope were made using the 1mm spectral
line receiver (A2), together with the Digital Autocorrelation Spectrometer
(DAS) in wide-band mode (750 MHz bandwidth).  Because of excess noise near
the edge of the passband, the usable bandwidth was only $\sim$700 MHz,
which corresponds to a total velocity coverage of $\sim$900 km s$^{-1}$ at
230 GHz.

All observations were obtained using a nutating subreflector with a chop
rate of $\sim$1.25 Hz.  Data were stored as six minute scans, and a
chopper wheel calibration was performed after every other scan.  Pointing
was monitored every few hours by observations of standard continuum
sources and was estimated to be accurate to within a few arcseconds.
During the course of the observations, data were also taken with the
bandpass centered at velocity offsets $\pm$ 40 km s$^{-1}$ from the
velocity corresponding to the redshift of the optical emission lines.
Shifting the velocities in such a manner minimizes any ripples inherent in
the baseline.

\subsubsection{Instituto de Radio Astronomia 
Milim\'{e}trica (IRAM) 30m Telescope}

The IRAM telescope has the ability to perform observations using the 3mm,
2mm, and 1mm receivers simultaneously, making use of the 2 available 512
MHz filterbanks and two 512 MHz autocorrelators.  For F09105+4108, the 2mm
and 1mm receivers were tuned to the redshifted CO($2\to1$) and CO($3\to2$)
transitions, respectively.  During the observations, pointing was
monitored by observing the planets and standard continuum sources and was
typically accurate to within 3 arcseconds.

\subsubsection{National Radio Astronomy Observatory (NRAO) 12m Telescope}

Observations of the redshifted CO($2\to1$) emission in PG{\ts}1634+706
were made with the NRAO 12m Telescope configured with a dual polarization
SIS spectral-line receiver and two $256\times2$ MHz channel filterbanks.
Observations were obtained using a nutating subreflector with a chop rate
of $\sim$1.25 Hz.  Six minute scans were taken, and a calibration was done
every other scan. The pointing was checked every few hours by observing
Jupiter and was estimated to be accurate to within a few arcseconds.
Because the duration of the observation was equivalent to one source
transit, no velocity shifts were made.

\section{Results}

\subsection{Imaging Data}

The I-band data reduction was done with IRAF.  The data reduction
consisted of flatfielding the individual images, scaling each image to its
median value, and shifting and averaging the images.

The K$^\prime$-band data reduction was done similar to the I-band data
reduction, except that after the median level of each individual frame was
subtracted, the individual frames were averaged together (without spatial
shifting) using a minmax averaging routine in order to create a ``sky''
image. This sky image was subtracted from the individual frames before
they were shifted and averaged.

Because the R-band image of F15307+3252 consisted of one integration, the
image was simply flatfielded and scaled to its median value. Cosmic rays
were then removed from the area around the galaxy. With the exception of
the K$^\prime$-band image of PG{\ts}1634+706, which shows only the
unresolved point-like QSO, the final reduced versions of all the images
are shown in Figures 1--3.

\subsection{Near-Infrared Spectra}

The KSPEC data reduction was done with IRAF.  The sky frames were first
subtracted from the source frames.  The sky subtracted frames were then
averaged and divided by the flat fields. Because there were only a few bad
pixels and none were on the spectral area containing the source spectrum,
the bad pixels were individually set to zero before the spectral orders
were extracted with the APALL package. The extracted spectra were then
wavelength calibrated and divided by the standard star spectrum, reduced
in the same fashion as the source spectrum, to remove any instrumental
effects and atmospheric lines (note that the stellar absorption lines in
the standard star were removed using the IRAF task SPLOT before the
division was done).  Because of the narrow width of the slit, no
photometry was determined for any of the KSPEC observations. Finally the
spectrum was multiplied by a Planck blackbody spectrum with the same
temperature as the standard star.

Table 5 summarizes the emission-line properties determined for F09105+4108
and F15307+3252.  The emission-line fluxes, $f(\lambda)$, line widths,
$\Delta v_{\rm FWHM}$, and root-mean-square flux densities, $S_{\rm rms}$,
were determined using the IRAF task SPLOT.  The signal-to-noise ratios
were determined by using the expression $f(\lambda) / \Delta v_{\rm FWHM}
S_{\rm rms}$.

The CGS4 data were reduced in a same manner as the KSPEC data, except that
the flat fielding and masking of bad pixels was done automatically after
every observation.

Figure 4 shows the spectra of F09105+4108 obtained with KSPEC.  Six of the
eight emission lines detected are at wavelengths corresponding to a source
redshift of $z = 0.4417$, identical to the observer-frame optical redshift
determined by Kleinmann et al. (1988). The discrepancy in the [O
III]{\ts}$\lambda \lambda$4959,5007 doublet redshift ($z = 0.4429$) is due
to wavelength calibration inaccuracies (the overlapping of the last few
orders made it difficult to determine which argon line was in which
order).

Figure 5 shows the spectra of F15307+3252 obtained with KSPEC and the
J-band spectrum obtained with CGS4. The KSPEC spectral lines imply a
source redshift of $z = 0.926$, and the CGS4 spectra imply a redshift of
$z = 0.927$, both consistent with the redshift of 0.926 determined by
Cutri et al. (1994).  Note that both F09105+4108 and F15307+3252 show
strong forbidden-line emission. Further, the H$\alpha$+[N II]
$\lambda\lambda6548+6583$ and [S II] $\lambda$6724 lines widths of
F09105+4108 (Table 5) are consistent with those measured from median
resolution spectroscopy by Liu, Graham, \& Wright (1996).  The
interpretation of the line emission will be discussed in \S 4.2.

To determine the fluxes of the H$\alpha$ and [N II]{\ts}$\lambda
\lambda$6548+6583 lines of F09105+4108 and F15307+3252, the line profiles
were fit using the IRAF package SPECFIT.  Initial guesses to the fluxes
and line widths were entered into the program, and the line flux ratios
and wavelength ratio of [N II]{\ts}$\lambda$6583 to [N
II]{\ts}$\lambda$6548 were fixed to 3 and 1.005345 (i.e., the ratio
$6583/6548$), respectively.  In addition, the line widths of the two [N
II] lines were fixed to be equal.  Iterations were made until a convincing
fit was achieved.  The best fit models are listed in Table 5.

Figure 6 is a short exposure of the redshifted [O
III]{\ts}$\lambda$$\lambda$4959,5007 and H$\alpha$ lines of
PG{\ts}1634+706. The redshifts determined from the $\Delta v_{\rm FWHM}$ =
4600 km s$^{-1}$ H$\alpha$ line and the [O III]{\ts}$\lambda$5007 lines
are $z = 1.339$ and 1.337, respectively.  The systemic redshift determined
by Tytler \& Fan (1992) is 1.3371.

\subsection{CO Spectra}

Table 6 summarizes the CO line data for the 3 sources.  In the case of the
IRAM 30m observations, where data was taken at several widely spaced
velocity offsets, the data are presented in separate spectral blocks.
Separating the data in this manner avoids the creation of direct current
(dc) offset-induced ghost features (see Evans et al. 1996) and provides
data blocks with relatively constant signal-to-noise across each
spectrum.

Figures 7--9 show the spectra of F09105+4108, F15307+3252, and
PG{\ts}1634+706, respectively.  For each source, individual scans were
examined to check for sinusoids in the baselines and other
irregularities.  After bad scans were removed, the rest of the scans were
averaged and linear baselines were subtracted.  For the JCMT 15m data,
where the sources were observed with several small velocity offsets, only
the overlapping regions of the scans were averaged.  Finally, the spectra
were smoothed to 30-50 km s$^{-1}$.  No emission lines were detected.
More sensitive upper limits of the CO emission in PG 1634+706 and
F15307+3252 have also been recently reported by Barvainis et al. (1998)
and Yun, Scoville, \& Evans (1998), respectively (see Table 6).

\subsubsection{CO Luminosity and H$_2$ Mass Limits}

Table 6 lists the root-mean-square temperatures, $T_{\rm rms}$, achieved
with a velocity resolution, $\Delta v _{\rm res}$, for the 3 sources
observed. To calculate the upper limit of the CO line intensity, $I_{\rm
CO}$, we assume that the CO lines have a FWHM velocity, $\Delta v _{\rm
FWHM} \sim 250$ km s$^{-1}$, similar to CO lines seen in local ULIGs (e.g.
Sanders et al. 1991) and in  IRAS{\ts}10214+4724 (e.g. Solomon et al.
1992a), Using the observed $T_{\rm rms}$ and $\Delta v_{\rm res}$ from
Table 6, the 3$\sigma$ limit on $I_{\rm CO}$ is given by 
$${I_{CO}} < {{3
T_{\rm rms} \Delta v_{\rm FWHM}} \over {\sqrt {\Delta v_{\rm FWHM} /
\Delta v_{\rm res}}}}\ ~[{\rm K~km~s}^{-1}]. \eqno(1)$$ 
Multiplying
$I_{\rm CO}$ by the Kelvin-to-Jansky conversion for a point
source\footnote{ Kelvin-to-Jansky conversion factors:\ JCMT 15m -- 15.6 Jy
K$^{-1}$ (Matthews 1992), IRAM 30m -- 4.6 to 4.9 Jy K$^{-1}$ (Guelin,
Kramer, \& Wild 1995), NRAO 12m -- 25.2 Jy K$^{-1}$ (P. Jewell, private
communication)} gives the corresponding limit for $S_{\rm CO} \Delta v$
[Jy km s$^{-1}$].

The luminosity distance for a
source at a given redshift, $z$, is,
$$D_{\rm L} = cH^{-1}_0 q^{-2}_0 \left\{ zq_0 + \left( q_0 - 1 \right)
\left( \sqrt {2q_0z + 1} - 1 \right) \right\} ~~~[h^{-1} {\rm ~Mpc}],
\eqno(2)$$
where $q_0$ is the deceleration parameter (= 0.5 for a critical
density universe) and $H_0 =
100 h$
(km s$^{-1}$ Mpc$^{-1}$).
Given the estimated upper limit of
$S_{\rm CO} \Delta v$, we
can calculate the CO luminosity of a source at
redshift $z$,
$$L'_{\rm CO} = \left( {c^2 \over {2 k \nu^2_{\rm obs}}} \right)
S_{\rm CO} \Delta v D^2_{\rm L} (1 + z)^{-3}, \eqno(3)$$
where $c$ is the speed of light, $k$ is the Boltzmann constant,
and $\nu_{\rm obs}$ is the observed frequency.
In terms of useful units, $L'_{\rm CO}$ for the CO($J\to{J-1}$)
rotational transition can be written as,
$$L'_{\rm CO} = 2.4\times10^3
\left( S_{\rm CO} \Delta v \over {\rm Jy~km~s}^{-1} \right)
\left( D^2_{\rm L} \over {\rm Mpc^2} \right) J^{-2} (1 + z)^{-1}
~~[h^{-2} {\rm K~km~s}^{-1} {\rm~pc}^2]. \eqno(4)$$

Table 6 lists the $L'_{\rm CO}$ upper limits of the 3 sources observed.
We have assumed that CO rotational transitions up to CO($4\to3$) have
approximately the same brightness temperature.  This assumption is based
on previously published multi-transition CO analyses of the HyLIG
IRAS{\ts}10214+4724 (e.g. Solomon et al.  1992a). For PG{\ts}1634+706, we
have assumed that the CO($5\to4$) transition has approximately the same
brightness temperature as the lower transitions, as observed in the
Cloverleaf Quasar (Barvainis 1996).

The upper limits on H$_2$ masses implied by the observed limits on
$L^\prime_{\rm CO}$ can be determined by making the reasonable assumption
that the CO emission is optically thick and thermalized, and that it
originates in gravitationally bound molecular clouds.  For molecular gas
in gravitationally bound clouds, the ratio of the H$_2$ mass and the CO
luminosity is given by $\alpha = M($H$_2) / L^\prime_{\rm CO} \propto
\sqrt {n({\rm H}_2)} / T_{\rm b}$ $M_\odot$ (K km s$^{-1}$ pc$^2)^{-1}$,
where $n($H$_2)$ and $T_{\rm b}$ are the density and brightness
temperature for the appropriate CO transition (e.g.  Sanders \& Scoville
1987; Solomon et al. 1992a).  Multi-transition CO surveys of molecular
clouds in the Milky Way (e.g. Sanders et al. 1993), and in nearby
starburst galaxies (e.g. G\"{u}sten et al. 1993) have shown that hotter
clouds tend to be denser such that the density and temperature
dependencies tend to cancel each other.  The variation in the value of
$\alpha$ is less than a factor of 2 for a wide range of gas kinetic
temperature, gas densities, and CO abundances.  We adopt a value of 4
$M_\odot$ (K km s$^{-1}$ pc$^2)^{-1}$ for $\alpha$, which is similar to
the value determined for the bulk of the molecular gas in the disk of the
Milky Way (c.f. Scoville \& Sanders 1987), as well as the value determined
from a multi-transition CO analysis of IRAS{\ts}10214+4724 (Solomon et al.
1992a).  Table 6 lists the upper limits on the the molecular gas mass of
F09105+4108, F15307+3252, and PG{\ts}1634+706 determined using this value
of $\alpha$.

\section{Discussion}

As stated in the introduction, spectroscopy has been a powerful tool for
determining the dominant ionization source in infrared-luminous galaxies.
The environment and morphology of the host galaxy are an essential
supplement to the spectroscopy, often providing information on global
spatial scales that have direct bearing on the nuclear activity.  We
briefly discuss the imaging data, then dedicate the bulk of the remaining
discussion to the spectroscopic data.

\subsection{Morphologies and Environment}

Figures 1-3 show images of the three sources taken at wavelengths longward
of V-band.  Most local ($z<0.3$) ultraluminous infrared galaxies are
observed to have distorted morphologies, multiple nuclei, and close
companions indicative of the merger of two or more galaxies (e.g. Sanders
\& Mirabel 1996 and references therein; Surace et al. 1997).  F09105+4108
clearly shows evidence for companions (Figure 1).  Indeed, unlike most
luminous infrared galaxies (LIGs), F09105+4108 is clearly a cD galaxy.
Both the I and K$^\prime$-band images of the galaxy show evidence of
``clumps'' within the envelope of the cD galaxy, which may be galaxies
stripped by tidal or dynamical friction.  Such stripping may be the source
of dust for F09105+4108.  In addition, F09105+4108 is known to have radio
jets (Hines \& Wills 1993), indicating the presence of a buried AGN.

While the R-band image of F15307+3252 (Figure 2a) shows no direct evidence
of distorted morphologies or close companions, the K$^{\prime}$-band image
Figure 2b) shows a faint companion to the southwest of the main galaxy.
Soifer et al. (1996) and Liu, Graham, \& Wright (1996) also present
near-infrared images showing this companion, with the latter group showing
evidence for a second companion. No evidence is seen for tidal features in
Figure 2, or in any other images published to date.

The QSO PG{\ts}1634+706 (Figure 3), the most distant of the three sources,
shows no evidence of distortions or close companions. There appear to be
approximately 30 faint objects in the vicinity of the QSO with $m_I =
23.5-25.5$, which is consistent with the density of faint field objects
expected within the given magnitude range (Pozzetti et al. 1998).  Thus,
there is also no apparent evidence with the present data set that the QSO
is imbedded in a cluster.

\subsection{Emission-Line Diagnostics}

Figure 10 shows 4 emission-line diagnostic diagrams commonly used to
distinguish galaxies with Seyfert/LINER-like properties from galaxies with
H II region-like properties.  The usefulness of these diagrams can be
understood by the following emission-line characteristics:  Emission from
[S II]{\ts}$\lambda$6724 and [N II]{\ts}$\lambda \lambda$6548+6583 can
emanate from ionized hydrogen regions, and [S II]{\ts}$\lambda$6724 and [O
I]{\ts}$\lambda$6300 emission can emanate from semi-ionized regions where
collisional ionization is significant.  Enhancement of these lines occurs
in AGNs because, unlike H II regions, they have extended partially ionized
zones created by an excess of X-rays (the absorption cross sections of
neutral hydrogen, helium and all ions are small for X-rays, thus X-rays
tend to escape the ionized region before interacting: Veilleux \&
Osterbrock 1987).  [O III]{\ts}$\lambda$5007 is a high ionization line
photoionized by UV photons, and thus tends to be strong in Seyfert
galaxies.  [S III]{\ts}$\lambda \lambda$9069+9532 is not as sensitive to
the level of ionization as [O III]{\ts}$\lambda$5007, most likely due to
its lower ionization potential (Osterbrock, Tran, \& Veilleux 1992).
However, [S III]{\ts}$\lambda \lambda$9069+9532 and [S
II]{\ts}$\lambda$6724 together do provide some degree of separation
between the 3 classes of active galaxies (Figure 10d).

Emission-line ratios of F09105+4108 and F15307+3252 are plotted in the
diagnostic diagrams.  For F09105+4108, the [O III]{\ts}$\lambda$5007/
H$\beta$ and [O I]{\ts}$\lambda$6300 / H$\alpha$ ratios have been
determined from Kleinmann et al. (1988).  Because their [O
III]{\ts}$\lambda$5007/ H$\beta$ ratio is being used in combination with
our KSPEC measurements, it is worth cautioning that [O
III]{\ts}$\lambda$5007/ H$\beta$ may vary as a function of the aperture
size used to observe F09105+4108.  In all 4 diagrams, both galaxies lie in
the Seyfert region, indicating that the dominant source of ionization is a
hard, nonthermal continuum source.  The analysis of these data are
consistent with the detection of broad emission lines in polarized light
in all three sources (Hines \& Wills 1993; Hines et al. 1995; Jannuzi et
al. 1994) and previous optical emission-line diagnostics of both galaxies
done with low resolution (Soifer et al. 1994, 1996) and median resolution
(Kleinmann et al. 1988; Liu, Graham, \& Wright 1996) spectroscopy.

Table 7 is a comparison of the observed [O III]{\ts}$\lambda$5007 /
H$\alpha$, [S II]{\ts}$\lambda$6724 / H$\alpha$, and [N
II]{\ts}$\lambda$6548+6583 / H$\alpha$ ratios of F09105+4108 and
F15307+3252 with the sample of Bright {\it IRAS} galaxies and warm {\it
IRAS} galaxies (which have been divided into Seyfert 2, LINER, and H II
region classifications). For F09105+4108, the ratio [O
III]{\ts}$\lambda$5007 / H$\alpha$ has been determined from Kleinmann et
al. (1988), who calculate [O III]{\ts}$\lambda$5007 / H$\beta \sim 12$ and
a negligible extinction (note that the [O III]{\ts}$\lambda$5007 /
H$\alpha$ ratio calculated from our data indicate that half of the [O
III]{\ts}$\lambda$5007 flux is lost via defocussing; see \S 2.2.1).
Despite the obvious scatter in the [O III]{\ts}$\lambda$5007 / H$\alpha$
ratio due to dust, there is a notable separation between the ratio for
Seyferts and for LINERs and H II region galaxies (i.e., the separation is
due mainly to the ionization parameter).  While F09105+4108 has a [O
III]{\ts}$\lambda$5007 / H$\alpha$ ratio 2 standard deviations above the
average ratio for low-redshift Seyfert 2 galaxies, F15307+3252 have
observed [O III]{\ts}$\lambda$5007 / H$\alpha$ ratio higher than the
average Seyfert 2 ratio, but still within the standard deviation of the
Seyfert 2 ratios. Inversely, the [S II]{\ts}$\lambda$6724 / H$\alpha$ and
[N II]{\ts}$\lambda$6548+6583 / H$\alpha$ ratios are lower than the
average values of the Seyferts, but still within the standard deviation of
the Seyfert 2 ratios. This may imply that the hyperluminous galaxies have
smaller semi-ionized regions or lower metallicities than the average,
local {\it IRAS} Seyferts.

\subsection{$L_{\rm ir}$ versus $L'_{\rm CO}$} 

Given the evidence presented here and elsewhere for AGNs in both
F09105+4108 and F15307+3252, we consider the infrared and CO luminosities
of these 2 sources and PG{\ts}1634+706 relative to LIGs, ULIGs and QSOs
detected by {\it IRAS}, but having lower infrared luminosities.  Figure
11a is a plot of $L'_{\rm CO}$ vs. $L_{\rm ir}$; the upper limits on the
$L'_{\rm CO}$ of F09105+4108, F15307+3252, and PG{\ts}1634+706 are less
than the $L'_{\rm CO}$ of the most gas-rich galaxies observed locally,
perhaps indicating that the upward trend of data points in the diagram
levels off at higher $L_{\rm ir}$, or disperses.  Figure 11b is a plot of
$L_{\rm ir}/L'_{\rm CO}$ vs. $L_{\rm ir}$; the lower limits on the $L_{\rm
ir} / L'_{\rm CO}$ of the 3 sources are consistent with the increasing
values of $L_{\rm ir} / L'_{\rm CO}$ as a function of increasing $L_{\rm
ir}$ of the sample of LIGs and ULIGs.  One possible explanation for the
data trends observed in Figure 11 is that $L_{\rm ir}$ becomes
increasingly dominated by reprocessed nonstellar light at high $L_{\rm
ir}$. From the spectral energy distributions plotted in
Figure 12, it is clear that all 3 sources have $\nu L_\nu(60 \mu{\rm m}) /
\nu L_\nu(100 \mu{\rm m})$ ratios higher than nearby, less luminous ULIGs
that have been detected in CO such as Arp 220 and Mrk 1014 (e.g. Sanders
et al. 1988b).  The mass of ``warm'' dust required to produce the observed
infrared luminosity can be computed by fitting the rest-frame far-infrared
(40-100 \micron) emission with a single-temperature dust model using dust
emissivities from Draine \& Lee (1984).  The dust temperatures implied by
the far-infrared SEDs in Figure 12 are in the range 100--200{\ts}K, and
the corresponding dust masses, $M_{\rm d} = (L_{\rm ir}/10^8{\ts}L_{\sun})
(40{\ts}{\rm K}/T_{\rm d})^5$, are in the range $10^7 -
10^5{\ts}M_{\sun}$.  Assuming, as is found for ULIGs like Arp 220 and Mrk
1014, that the bulk of the dust is heated to the computed ``warm'' dust
temperature, and that the gas/dust ratio is similar to what is found for
other ULIGs such as Arp 220 (i.e. $\sim${\ts}200: Sanders, Scoville, \&
Soifer 1991), then the implied range of H$_2$ gas mass is $10^{9.3} -
10^{7.3}{\ts}M_{\sun}$.  Thus it is possible that the observed infrared
luminosities are produced by a relatively modest mass of gas and dust
(e.g. comparable to or less than the total molecular gas mass of the Milky
Way) heated to temperatures 100--200{\ts}K, perhaps by the central AGN.
Such small amounts of molecular gas would also explain our non detections
of all three sources in CO (see Table 6).  Further, while it is clear from
the observed narrow emission lines and large infrared excess ($L_{\rm
ir}/L_{\rm bol} > 0.9$) of F09105+4108 and F15307+3252 that the
circumnuclear covering factor of the dust is almost unity, the broad
emission lines and smaller infrared excess ($L_{\rm ir}/L_{\rm bol} \sim
0.35$) of PG{\ts}1634+706 indicate that the QSO has a low covering
factor.

It is curious that CO surveys of galaxies at cosmological distances have
yielded no evidence of galaxies with $L'_{\rm CO}$ much larger than that
of the most CO luminous galaxies at lower redshift ($z \lesssim 0.2$)
plotted in Figure 11 (i.e., $L'_{\rm CO}$ corresponding to  $\sim$
3--4$\times 10^{10}{\ts}h^{-2}{\ts}M_\odot$ of molecular gas).  To date,
there exists little evidence that the H$_2$ content of active galaxies
evolves with redshift ($z \lesssim 4$).  The extremely high ${\rm H}_2$
masses originally computed for IRAS{\ts}10214+4724 and H{\ts}1413+117 were
in large part due to gravitational lensing of these sources by foreground
objects (IRAS{\ts}10214+4724: Elston et al. 1994; Soifer et al. 1995;
Trentham 1995; Graham \& Liu 1995; Broadhurst \& Leh\'{a}r 1995; Serjeant
et al.  1995; Close et al. 1995; Eisenhardt et al. 1995; Downes, Solomon,
\& Radford 1995; H{\ts}1413+117: Barvainis et al. 1994), and CO surveys of
$1<z<4$ high redshift powerful radio galaxies (HzPRGs: Evans et al. 1996;
Downes et al. 1996) and $z\sim4$ luminous high redshift QSOs (Barvainis \&
Antonucci 1996) were unsuccessful at detecting sources with more molecular
gas than the most gas-rich galaxies observed locally (i.e. based on
3$\sigma$ upper $M({\rm H}_2)$ limits of the HzPRGs and high redshift
QSOs).  The reported detection of CO emission in the $z = 4.7$ QSO
BR{\ts}1202-07 ($M({\rm H}_2) \sim 6\times10^{10}{\ts}h^{-2}{\ts}M_\odot$)
may be the first indication that $M({\rm H}_2)$ increases beyond $z \sim
4$. However, reports of shear due to gravitational lensing in the vicinity
of BR{\ts}1202-07 may mean that the CO luminosity is being amplified (e.g.
see discussion in Omont et al. 1996b).  Further, there have been recent,
unconfirmed detections of CO emission in the $z = 2.4$ radio galaxy 53W002
(Scoville et al. 1997), the $z = 4.5$ QSO BR{\ts}1335-07 (Guilloteau et
al. 1998), and the gravitationally lensed, $z=2.6$ quasar MG 0414+0534
(Barvainis et al.  1998), but all have molecular gas masses comparable to
the most gas-rich, local galaxies.  \footnote{This statement assumes that
the observed CO emission in MG 0414+0534 is gravitationally lensed by at
least three times its intrinsic luminosity.} The inferred molecular gas
masses and upper limits of high redshift galaxies surveyed to date may be
evidence that either galaxies with more than $\sim$ 3--4$\times
10^{10}{\ts}h^{-2}{\ts}M_\odot$ of molecular gas rapidly turn the gas into
stars, and/or that galaxies form in a hierarchical fashion.  It may also
be the case that only high-redshift galaxies with a significant nonstellar
contribution to their infrared luminosity will be observed to have high
$L_{\rm ir}$.  A better understanding of the relationship between extreme
$L_{\rm ir}$ and H$_2$ mass awaits the detection and observation of a
larger sample of HyLIGs, especially those having relatively cool {\it
IRAS} colors, i.e. relatively low values of $\nu L_\nu (60 \mu{\rm m}) /
\nu L_\nu (100 \mu{\rm m})$.

\section{Summary}

We have presented new millimeterwave CO(J+1,J) observations and rest-frame
0.48--1.1{\ts}$\mu$m spectroscopy of the HyLIGs IRAS{\ts}F09105+4108,
IRAS{\ts}F15307+3252, and PG{\ts}1634+706.  The following conclusions are
drawn:

({\it 1}) The emission line ratios 
[O III]{\ts}$\lambda$5007/H$\beta$, [O I]{\ts}$\lambda$6300/H$\alpha$, 
[N II]{\ts}$\lambda$6583/H$\alpha$,
[S II]{\ts}$\lambda$6724/H$\alpha$,
[S III]{\ts}$\lambda \lambda$9069+9532/H$\alpha$
of F09105+4108 and F15307+3252 indicate that both 
have Seyfert 2 emission-line spectra, consistent with previously
published work.

({\it 2}) Our upper limits on the molecular
gas mass of the 3 HyLIGs are comparable to or less than the
H$_2$ mass of the most gas-rich galaxies observed locally.

({\it 3}) The relatively ``warm'' far-infrared
(60{\ts}\micron/100{\ts}\micron) colors (implying dust temperatures
$T_{\rm d} = 100-200${\ts}K) for all three sources are consistent with the
idea that the bulk of their infrared emission emanates from a relatively
modest amount of warm circumnuclear dust, i.e. $M_{\rm d} = 10^5 -
10^7{\ts}M_{\sun}$, corresponding to a molecular gas mass, $M({\rm H}_2)
\simeq 10^{7.3}-10^{9.3}{\ts}M_{\sun}$.  This would be consistent with our
measured upper limits for $L'_{\rm CO}$ and the corresponding extreme
values of $L_{\rm ir} / L'_{\rm CO}$.

({\it 4}) 
While the narrow emission lines and $L_{\rm ir}/L_{\rm bol}
> 0.90$ of F09105+4108 and F15307+3252 indicate a
circumnuclear dust covering factor of almost unity,
the broad emission lines and $L_{\rm ir}/L_{\rm bol} \sim 0.35$
of PG{\ts}1634+706 indicate that most of the AGN is exposed.

\acknowledgements

We thank the staffs of the JCMT 15m telescope, the IRAM 30m telescope, the
NRAO 12m telescope, the UH 2.2m telescope, and the United Kingdom Infrared
Telescope for their generous support during our observations.  A.S.E.
thanks N. Trentham, E. Egami, J. Hora, D. Kim, T. Greene, P. Hall, and J.
Goldader for useful discussions and assistance during the preparation of
this paper.  A.S.E. is also indebted to J. Jensen and D. Jewitt for
obtaining part of the K$^{\prime}$-band and the R-band image data,
respectively, of F15307+3252, and to J. Mazzarella for help in obtaining
the IRAS SCANPI flux measurements for all three sources.  We also thank K.
Teramura for the final preparation of figures show in this paper, and the
referee, Sylvain Veilleux, for many useful comments.  A.S.E. was supported
in part by NASA grant NAG5-3042.  D.B.S. was supported in part by NASA
grant NAG5-3370.  This research has made use of the NASA/IPAC
Extragalactic Database (NED) which is operated by the Jet Propulsion
Laboratory.

\clearpage

\begin{deluxetable}{lllrrrrccc}
\scriptsize
\tablenum{1}
\tablewidth{0pt}
\tablecaption{Source Parameters}
\tablehead{
\multicolumn{1}{c}{Name} &
\multicolumn{1}{c}{R.A.} &
\multicolumn{1}{c}{Dec.} &
\multicolumn{4}{c}{$f_\nu(\lambda)$\ (Jy)} &
\multicolumn{1}{c}{{\it z}} &
\multicolumn{1}{c}{Refs} &
\multicolumn{1}{c}{$L_{\rm ir}$\tablenotemark{a}}\nl
\multicolumn{1}{c}{IRAS} & 
\multicolumn{2}{c}{B1950.0} & 
\multicolumn{1}{c}{12\micron} & 
\multicolumn{1}{c}{25\micron} & 
\multicolumn{1}{c}{60\micron} & 
\multicolumn{1}{c}{100\micron} & 
\multicolumn{2}{c}{} & 
\multicolumn{1}{c}{($L_{\sun}$)}\nl} 
\startdata
F09105+4108\tablenotemark{b}&09:10:32.84&41:08:53.6&0.13$\pm$.03&0.33$\pm$.03&0.53$\pm$.04&$<$0.32&0.4417&1,3,4&$7.8\times10^{12}h^{-2}$ \nl 
F15307+3252&15:30:44.63&32:52:50.6&$<$0.07&0.08$\pm$.07&0.26$\pm$.08&0.46$\pm$0.19&0.926&1,3,4&$2.5\times10^{13}h^{-2}$ \nl
F16348+7037{\ts}(PG~1634+706)&16:34:51.0&70:37:37.0&0.06$\pm$.01&0.13$\pm$.01&0.25$\pm$.03&$<$0.45&1.337&2,3,5&$5.7\times10^{13}h^{-2}$ \nl
\enddata
\tablenotetext{a}{Assuming $q_o = 0.5$, $H_{\rm o} = 100 h${\ts}km s$^{-1}${\ts}Mpc$^{-1}$.}
\tablenotetext{b}{IRAS Point Source Catalog name in Kleinmann et al. (1988) was P09104+4109}
\tablerefs{Positions, fluxes, and redshifts:\ (1) Becker et al. (1995);  (2) This paper (IRAS FSC); 
 (3) This paper (IRAS SCANPI); (4) This paper (near-infrared emission lines); (5)  Titler \& Fan (1992)}
\end{deluxetable}

\begin{deluxetable}{llccccc}
\tablenum{2}
\tablewidth{0pt}
\tablecaption{Journal of Imaging Observations}
\tablehead{
\colhead{Source} &
\multicolumn{1}{c}{Telescope} &
\multicolumn{1}{c}{Detector} &
\multicolumn{1}{c}{Pixel Scale\tablenotemark{a}} &
\multicolumn{1}{c}{Band} &
\multicolumn{1}{c}{Date} &
\colhead{Time}\nl
\colhead{} & 
\multicolumn{1}{c}{} &
\multicolumn{1}{c}{} &
\multicolumn{1}{c}{arcsec/pixel} &
\multicolumn{1}{c}{} &
\multicolumn{1}{c}{m/y} &
\colhead{min.}
} 
\startdata
IRAS F09105+4108&UH 2.2m&Tek&0.14&I&03/97&26 \nl
             &&QUIRC&0.19&K$^{\prime}$&04/96&30 \nl
IRAS F15307+3252&UH 2.2m&Tek&0.22&R&04/94&20 \nl
             &&QUIRC&0.19&K$^{\prime}$&04/96& 60 \nl
PG 1634+706&UH 2.2m&Orbit&0.14&I&06/97&55 \nl
           &&QUIRC&0.06&K$^{\prime}$&05/97&16 \nl
\tablenotetext{a}{Imaging at the UH 2.2m Telescope is done with a
focal ratio of f/10 or 
f/31. For the CCD observations, 
images taken at f/31 are binned 2$\times$2
pixels.}
\enddata
\end{deluxetable}

\begin{deluxetable}{lcccccc}
\tablenum{3}
\tablewidth{0pt}
\tablecaption{Journal of Near-Infrared Spectroscopic Observations}
\tablehead{
\colhead{Name} &
\multicolumn{1}{c}{Telescope} &
\multicolumn{1}{c}{Detector} & 
\multicolumn{1}{c}{Slit Width} &
\multicolumn{1}{c}{$\frac{\lambda}{\Delta \lambda}$} &
\multicolumn{1}{c}{Date} & 
\multicolumn{1}{c}{Time} \nl 
\colhead{} & 
\multicolumn{1}{c}{} & 
\multicolumn{1}{c}{} & 
\multicolumn{1}{c}{\arcsec} &
\multicolumn{1}{c}{} & 
\multicolumn{1}{c}{mm/yy} & 
\multicolumn{1}{c}{min} \nl 
}  
\startdata
IRAS~F09105+4108&UH 2.2m & KSPEC&1.0 & 620 & 12$/$94 & 18 \nl 
IRAS~F15307+3252&UH 2.2m & KSPEC & 0.8 & 760 & 03$/$94 & 33 \nl
                &UKIRT & CGS4 & 1.2 & 1000 & 05$/$95 & 56 \nl
PG~1634+706&UH 2.2m & KSPEC & 1.0 & 760 & 03$/$94 & 3 \nl
\enddata
\end{deluxetable}

\begin{deluxetable}{lrrrrc}
\tablenum{4}     
\tablewidth{0pt}
\tablecaption{Journal of CO Observations}
\tablehead{
\colhead{Name} & 
\multicolumn{1}{c}{line} &
\multicolumn{1}{c}{Telescope} &
\multicolumn{1}{c}{Date} & 
\multicolumn{1}{c}{Time} &
\colhead{T$_{\rm sys}$}\nl
\colhead{} &
\multicolumn{1}{c}{} & 
\multicolumn{1}{c}{} &
\multicolumn{1}{c}{m/y} & 
\multicolumn{1}{c}{hr} &
\colhead{K}\nl} 
\startdata
IRAS~F09105+4108&CO($3\to2$)&JCMT~15m&04$/$94&7.3&429\nl
 &CO($3\to2$)&IRAM~30m&06$/$94&6.3&724\nl
 &CO($2\to1$)&IRAM~30m&06$/$94&6.3&530\nl
IRAS~F15307+3252&CO($4\to3$)&JCMT~15m&03$/$94&8.4&380\nl
 &CO($4\to3$)&JCMT~15m&04$/$94&1.7&400\nl
 &CO($4\to3$)&IRAM~30m&06$/$94&10.2&590\nl
PG~1634+706&CO($5\to4$)&JCMT~15m&04$/$94&11.0&452\nl
 &CO($2\to1$)&NRAO~12m&05$/$94&5.6&160\nl
\enddata
\end{deluxetable}

\begin{deluxetable}{lcrrrrrrrr}
\tablenum{5}       
\scriptsize
\tablewidth{0pt}
\tablecaption{Near-Infrared Spectroscopy; Emission Line Properties}
\tablehead{
\colhead{} &
\colhead{} &
\multicolumn{4}{c}{IRAS~F09105+4108} &
\multicolumn{4}{c}{IRAS~F15307+3252} \nl
\nl
\cline{3-6} \cline{7-10} \\
\colhead{Line} &
\colhead{Instr.} &
\colhead{$\lambda _{\rm obs}$} & 
\colhead{$f(\lambda) / f({\rm [SII]})$} &
\colhead{FWHM\tablenotemark{a}} &
\colhead{$S/N~~~~$} &
\colhead{$\lambda _{\rm obs}$} &
\colhead{$f(\lambda) / f({\rm [SII]})$} &
\colhead{FWHM\tablenotemark{a}} &
\colhead{$S/N$} \nl
\colhead {} &
\colhead {} &
\colhead{(\AA)} &
\colhead{} &
\colhead{(km s$^{-1}$)} &
\colhead{} &
\colhead{(\AA)} &
\colhead{} &
\colhead{(km s$^{-1}$)} &
\colhead{} 
}
\startdata
H$\beta$ & KSPEC & \nodata & \nodata & \nodata & \nodata$~~~~$ & \nodata & $<1.4$ & (1300) & \nodata \nl
[O III]$\lambda$4959 & KSPEC & 7153.1 & 2.6$\pm$0.8 & 740 & 3.2$~~~~$ & 9556.5 & 5.5$\pm$0.4 & 1540 & 12 \nl
[O III]$\lambda$5007 & KSPEC & 7224.6 & 9.0$\pm$1.0 & 980 & 8.9$~~~~$ & 9644.6 & 16$\pm$0.4 & 1280 & 35 \nl
[O I]$\lambda$6300 & KSPEC & \nodata & $<0.44$ & (1000) & \nodata$~~~~$ & \nodata & $<1.0$ & (1300) & \nodata \nl
H$\alpha$ + [N II] & KSPEC & 9463.5\tablenotemark{b} & 6.8$\pm$0.2 & 1710 & 32$~~~~$ & 12641.8 & 9.8$\pm$0.5 & 1990 & 18 \nl
 & & 9490.6\tablenotemark{c} & & & & & & & \nl
\nl
[N II]$\lambda$6548\tablenotemark{d} & & 9441.7$\pm$2.9 & 0.62$\pm$0.09 & 1160$\pm$150 & \nodata$~~~~$  & 12618$\pm$5 & 1.4$\pm$0.2 & 1310$\pm$150 & \nodata \nl
H$\alpha$$\lambda$6563\tablenotemark{d} & & 9460.9$\pm$1.9 & 4.4$\pm$0.4 & 1200$\pm$100 & \nodata$~~~~$  & 12636$\pm$3 & 5.2$\pm$0.5 & 1300$\pm$150 & \nodata \nl
[N II]$\lambda$6583\tablenotemark{d} & & 9492.2$\pm$2.9 & 1.8$\pm$0.3 & 1160$\pm$150 & \nodata$~~~~$ & 12686$\pm$5 & 4.1$\pm$0.4 & 1310$\pm$150 & \nodata \nl
\nl
[S II]$\lambda$6724 & KSPEC & 9693.9 & 1.0$\pm$0.2 & 1120 & 6.2$~~~~$ & 12960.3 & 1.0$\pm$0.2 & 850 & 4.4 \nl
[S III]$\lambda$9069 & KSPEC & 13076.9 & 0.74$\pm$0.08 & 900 & 8.7$~~~~$ & 17467.1 & 1.1$\pm$0.3 & 630 & 3.5 \nl
[S III]$\lambda$9532 & KSPEC & 13174.3 & 2.0$\pm$0.08 & 930 & 20$~~~~$ & \nodata & \nodata & \nodata & \nodata \nl
He~I$\lambda$10830 & KSPEC & 15615.5 & 2.1$\pm$0.08 & 1070 & 27$~~~~$ & 20873.8 & 4.8$\pm$0.3 & 1540 & 16 \nl
\nl
[O I]$\lambda$6300 & CGS4 & \nodata & \nodata & \nodata & \nodata$~~~~$ & 12142.8 & $<1.6$ & (1300) & \nodata \nl
H$\alpha$ + [N II] & CGS4 & \nodata & \nodata & \nodata & \nodata$~~~~$ & 12650.4 & 7.1$\pm$0.7 & 1860 & 10 \nl
[S II]$\lambda$6724 & CGS4 & \nodata & \nodata & \nodata & \nodata$~~~~$ & 12959.6 & 1.0$\pm$0.3 & 1130 & 3.2 \nl
\enddata
\tablenotetext{a}{Uncorrected for instrumental broadening.}
\tablenotetext{b}{Observed wavelength of H$\alpha$$\lambda$6563.}
\tablenotetext{c}{Observed wavelength of [N II]$\lambda$6583.}
\tablenotetext{d}{Emission line values as determined by model fit.}
\tablecomments{The parenthetical velocities are adopted values based on the
average FWHM of the other, stronger emission lines.}
\end{deluxetable}

\begin{deluxetable}{lcrrrcrrrr}
\tablenum{6}         
\scriptsize
\tablewidth{0 pt}
\tablecaption{CO Emission Line Data}
\tablehead{
\colhead{Source} & \multicolumn{1}{c}{Site} & 
\multicolumn{1}{c}{line} &
\multicolumn{1}{c}{$\nu _{{\rm v}=0}$} & 
\multicolumn{1}{c}{v$_{\rm offset}$\tablenotemark{a}} &
\multicolumn{1}{c}{$\Delta {\rm v}_{\rm res}$} &
\multicolumn{1}{c}{$T_{\rm rms}$\tablenotemark{b}} &
\multicolumn{1}{c}{$S_{\rm CO}\Delta {\rm v}$} &
\colhead{$L'_{\rm CO}$\tablenotemark{c}} &
\colhead{$M$(H$_2$)\tablenotemark{d}}\nl
\colhead{} &
\multicolumn{1}{c}{} & 
\multicolumn{1}{c}{} & \multicolumn{1}{c}{(GHz)} &
\multicolumn{1}{c}{(km s$^{-1}$)} &
\multicolumn{1}{c}{(km s$^{-1}$)} & \multicolumn{1}{c}{(mK)} &
\multicolumn{1}{c}{(Jy km s$^{-1}$)} & 
\multicolumn{1}{l}{($\times 10^{9}h^{-2}$} &
\colhead{($\times 10^{10}h^{-2}$}\nl 
&
\colhead{} &
\multicolumn{1}{c}{} &
\multicolumn{1}{c}{} &
\multicolumn{1}{c}{} &
\multicolumn{1}{c}{} &
\multicolumn{1}{c}{} &
\multicolumn{1}{c}{} &
\multicolumn{1}{l}{ K km s$^{-1}$ pc$^2$)} &
\multicolumn{1}{r}{M$_\odot)$}\nl}
\startdata
IRAS F09105+4108&15m&CO($3\to2$)&239.853&-104&37&1.7&$<$7.7$~~~~$&$<$3.0$~~~~~~$&$<$1.2$~~~~~~$\nl
 &30m&CO($3\to2$)&239.853&0&31&6.2&$<$8.2$~~~~$&$<$3.1$~~~~~~$&$<$1.2$~~~~~~$\nl
 & & & &625&30&3.3&$<$4.3$~~~~$&$<$1.6$~~~~~~$&$<$0.66$~~~~~~$\nl
 &30m&CO($2\to1$)&159.907&0&38&4.6&$<$6.4$~~~~$&$<$5.6$~~~~~~$&$<$2.2$~~~~~~$\nl
 & & & &937&38&3.3&$<$4.7$~~~~$&$<$4.1$~~~~~~$&$<$1.6$~~~~~~$\nl
IRAS F15307+3252&15m&CO($4\to3$)&239.415&-108&38&1.2&$<$5.6$~~~~$&$<$4.5$~~~~~~$&$<$1.8$~~~~~~$\nl
 &30m&CO($4\to3$)&239.415&0&30&5.7&$<$7.4$~~~~$&$<$6.0$~~~~~~$&$<$2.4\tablenotemark{e}$~~~~~~$ \nl
 & & & &188&30&8.4&$<$11$~~~~$&$<$8.9$~~~~~~$&$<$3.5$~~~~~~$\nl
 & & & &-188&30&8.4&$<$11$~~~~$&$<$8.8$~~~~~~$&$<$3.5$~~~~~~$\nl
 & & & &814&30&3.6&$<$4.7$~~~~$&$<$3.8$~~~~~~$&$<$1.5$~~~~~~$\nl
 & & & &-814&30&3.0&$<$3.9$~~~~$&$<$3.1$~~~~~~$&$<$1.2$~~~~~~$\nl
PG 1634+706&12m&CO($2\to1$)&98.647&0&49&0.46&$<$3.8$~~~~$&$<$23$~~~~~~$&$<$9.2\tablenotemark{f}$~~~~~~$\nl
PG 1634+706&15m&CO($5\to4$)&246.584&0&36&1.6&$<$7.4$~~~~$&$<$7.1$~~~~~~$&$<$2.8$~~~~~~$\nl
\tablenotetext{a}{The difference between the velocity the receivers were
tuned to and the
the systemic velocity of the source.}
\tablenotetext{b}{The rms temperatures are given 
in terms of main beam brightness temperature.}
\tablenotetext{c}{Assuming $q_0 = 0.5$, $H_0 = 100$  $h$ km s$^{-1}$
Mpc$^{-1}$.}
\tablenotetext{d}{Assuming 
$\alpha = 4$~M$_\odot$ (K km s$^{-1}$ pc$^2)^{-1}$
(see text).}
\tablenotetext{e}{The 3$\sigma$ upper limit of M(H$_2$) reported by
Yun, Scoville, \& Evans (1998) over the redshift range 0.9240-0.09275
is $9.8\times10^{9}h^{-2}$ M$_\odot$.}
\tablenotetext{f}{The 3$\sigma$ upper limit of M(H$_2$) reported by
Barvainis et al. (1998) is $2.0\times10^{10}h^{-2}$ M$_\odot$.} 
\enddata
\end{deluxetable}

\begin{deluxetable}{lcccc}
\tablecolumns{5}
\tablenum{7}         
\tablewidth{0pt}
\tablecaption{Observed Emission-Line Ratios} 
\tablehead{
\colhead{Source} &
\multicolumn{1}{c}{${{\rm [O III]}\lambda5007 \over {\rm H}\alpha}$} &
\multicolumn{1}{c}{${{\rm [O III]} \over {\rm H}\beta}$} &
\multicolumn{1}{c}{${{\rm [N II]}\lambda \lambda 6548+6583 \over {\rm H}\alpha}$} &
\multicolumn{1}{c}{${{\rm [S II]}\lambda6724 \over {\rm H}\alpha}$} }
\startdata
\cutinhead{HyLIGs}
IRAS F09105+4108 & $3.9$\tablenotemark{a} & \nodata & $0.55\pm0.2$ & $0.23\pm0.11$ \nl
IRAS F15307+3252 & $3.0\pm0.4$ & $>11.1\pm0.3$ & $1.1\pm0.2$ & $0.19\pm0.10$ \nl
\cutinhead{{\it IRAS} Bright Galaxies}
Seyfert2 & $1.8\pm1.0$ & $10.2\pm4.6$ & $1.3\pm0.7$ & $0.41\pm0.26$ \nl
LINERs   & $0.16\pm0.10$ & $1.7\pm1.0$ & $1.7\pm1.1$ & $0.69\pm0.41$ \nl
H II     & $0.10\pm0.11$ & $0.64\pm0.52$ & $0.62\pm0.24$ & $0.27\pm0.06$ \nl
\enddata
\tablenotetext{a}{Determined from [O III] $\lambda$5007 / H$\beta = 12$
(Kleinmann et al. 1988)
and H$\alpha$ / H$\beta \sim 3.1$. The H$\alpha$ / H$\beta$ ratio is
based on the determination by  Kleinmann et al (1988)
that $E(B-V) < 0.15$ for F09105+4108.}
\end{deluxetable}

\vfill\eject
 
\centerline{Figure Captions}
 
\vskip 0.3in
 
\noindent
Figure 1. Broad-band images of the IRAS{\ts}F09105+4108;
a \& b) I-band images with $56\arcsec\times56\arcsec$
and $\sim 180\arcsec\times180\arcsec$ fields of view,
respectively. 
c \& d) K$^\prime$-band images with $56\arcsec\times56\arcsec$
and $143\arcsec\times143\arcsec$ fields of view, respectively.
For all images, North is up and E is to the left.

\noindent
Figure 2.  Broad-band images of the IRAS{\ts}F15307+3252;
a \& b) R-band images with $60\arcsec\times60\arcsec$
and $\sim 180\arcsec\times180\arcsec$ fields of view,
respectively.
c \& d) K$^\prime$-band images with $60\arcsec\times60\arcsec$
and $180\arcsec\times180\arcsec$ fields of view, respectively.
For all images, North is up and E is to the left.

\noindent
Figure 3. I-band Image of PG{\ts}1634+706 with a
$72\arcsec\times72\arcsec$ field of view.

\noindent
Figure 4.  KSPEC spectroscopy of IRAS{\ts}F09105+4108 covering the
observer-frame wavelength range 7000--17500 \AA.  The pixel sampling for
the R, I2, J1, and H-bands are 7.4, 8.5, 12, and 14 $\mu$m pixel$^{-1}$,
respectively.

\noindent
Figure 5.
 KSPEC spectroscopy of IRAS{\ts}F15307+3252 covering the observer-frame
wavelength range 8600--22400 \AA. Note that the feature in panel (a) at
17620 \AA$~$ is an OH sky line.  The pixel sampling for the I2, J2, H, and
K$^\prime$-bands are 8.5, 12, 14, and 20 $\mu$m pixel$^{-1}$,
respectively.  Panel (e) is a CGS4 spectrum of the H$\alpha$ + [N II] and
[S II]{\ts}$\lambda$6724 lines.  The pixel sampling for the CGS4 spectrum
is 4.4 $\mu$m pixel$^{-1}$.

\noindent
Figure 6.  KSPEC spectrum of [O III]{\ts}$\lambda$5007 and H$\alpha$
emission from PG{\ts}1634+706.  The pixel sampling for the H and
K$^\prime$-bands are 12 and 14  $\mu$m pixel$^{-1}$, respectively.  Note
that the broad Fe II complex is also evident redward of 12000\AA.

\noindent
Figure 7.  JCMT 15m and IRAM 30m spectra of IRAS{\ts}F09105+4108 ($z =
0.4417$).  The intensity scale is main-beam brightness temperature (IRAM)
and antenna temperature corrected for aperture losses (JCMT).  A linear
baseline has been subtracted from each spectrum.  The zero velocity
corresponds to the average redshift of optical/near-infrared emission
lines observed for IRAS{\ts}F09105+4108.

\noindent
Figure 8.  JCMT 15m and IRAM 30m spectra of IRAS{\ts}F15307+3252 ($z =
0.926$).  The intensity scale is main beam brightness temperature (IRAM)
and antenna temperature corrected for aperture losses (JCMT).  A linear
baseline has been subtracted from each spectrum.  The zero velocity
corresponds to the average redshift of optical/near-infrared emission
lines observed for IRAS{\ts}F15307+3252.

\noindent
Figure 9.
NRAO 12m and JCMT 15m spectra of PG{\ts}1634+706 ($z = 1.338$).
The intensity scale is main beam brightness temperature (NRAO)
and antenna temperature corrected for aperture losses (JCMT).
A linear baseline has been subtracted from each spectrum.
The zero velocity corresponds to the systemic velocity of the QSO
as determined by Tytler \& Fan (1992)
averaged with the redshift of the [O III]{\ts}$\lambda$5007
and H$\alpha$ emission lines.

\noindent
Figure 10.
Four emission-line diagnostic diagram plots of IRAS{\ts}F09105+4108
and IRAS{\ts}F15307+3252. The [O I] $\lambda$6300 / H$\alpha$ ratio
for F09105+4108
has been determined from the data in Kleinmann et al. (1988), 
where the H$\alpha$
flux was determined from the H$\beta$ flux and the observation that
the extinction in F09105+4108 is neglegible. 
Additional data of H II regions, LINERS, 
and Seyfert are taken from data compiled by Osterbrock et al. (1992).

\noindent
Figure 11.
(a). $L'_{\rm CO}$ vs. $L_{\rm ir}$ plot of {\it IRAS} galaxies and
QSOs.  Arrows denote upper limits.
References for CO and infrared (ir) data:
LIGs - Sanders, Scoville, \& Soifer (1991);
ULIGs - Sanders (1991);
IRAS{\ts}F09105+4108 - This work (CO) and Kleinmann et al. (1988) (ir);
IRAS{\ts}F15307+3252 - This work (CO) and Cutri et al (1994) (ir);
I{\ts}Zw{\ts}1 - Barvainis, Alloin, \& Antonucci (1989);
Mrk{\ts}1014 - Sanders, Scoville, \& Soifer (1988a);
3C{\ts}48 -  Scoville et al. (1993);
PG{\ts}1634+706 - This work (CO) and Sanders et al. (1989) (ir).
(b). $L_{\rm ir} / L'_{\rm CO}$
vs. $L_{\rm ir}$ plot of the same sources.
Arrows denote upper limits.

\noindent
Figure 12.
Rest-frame SEDs of IRAS{\ts}F09105+4108, IRAS{\ts}F15307+3252, and 
 PG{\ts}1634+706.
The dashed lines represent smooth fits to the data.
References for data:
For IRAS{\ts}F09105+4108 - all data are from Kleinmann et al. 1988.
For IRAS{\ts}F15307+3252 - all data are from Cutri et al. 1994.
For PG{\ts}1634+706 - all data from Sanders et al. (1989), except for
the 560 $\micron$ data (Chini, Kreysa, \& Bierman 1989) and data
between $10^3$--$10^5 \mu$m (Antonucci \& Barvainis 1988).


\begin{thebibliography}{}
\bibitem[]{} 
Antonucci, R. \& Barvainis, R. 1988, \apj, 332, L13

\bibitem[]{}
Barvainis, R. 1996, in CO: Twenty-five Years of Millimeter-wave
Spectroscopy, eds. W. B. Latter et al. (Dordrecht: Kluwer), 335 

\bibitem[]{}
Barvainis, R., Alloin, D., \& Antonnuci, R. 1989, \apjl, 337, L69

\bibitem[]{}
Barvainis, R., Alloin, D., Guilloteau, S., Antonucci, R. 1998,
\apj, 492, L13

\bibitem[]{}
Barvainis, R. \& Antonucci, R. 1996, \pasp, 108, 187

\bibitem[]{}
Barvainis, R., Tacconi, L., Antonucci, R., Alloin, D., \& Coleman, P.
1994, Nature, 371, 586

\bibitem[]{}
Broadhurst, T. \& Leh\'{a}r, J. 1995, \apjl, 450, L41 

\bibitem[]{}
Brown, R. L., \& Vanden Bout, P. A. 1991, \aj, 102, 1956

\bibitem[]{}
Brown, R. L., \& Vanden Bout, P. A. 1992, \apj, 397, L19

\bibitem[]{}
Chini, R., Kreysa, E., Bierman, P. L. 1989, A\&A, 219, 87

\bibitem[]{}
Close, L. M., Hall, P. B., Liu, C. T., \& Hege, E. K. 1995,
\apjl, 452, L9 

\bibitem[]{}
Cutri, R. M., Huchra, J. P., Low, F. J., Brown, R. L.,
\& Vanden Bout, P. A. 1994, \apjl, 424, L65

\bibitem[]{}
Downes, D., Solomon, P. M., \& Radford, S. J. E. 1995, \apj, 453, L65 

\bibitem[]{}
Downes, D. Solomon, P. M., Sanders, D. B., \& Evans, A. S. 1996, A\&A,
313, 91

\bibitem[]{}
Draine, B. T., \& Lee, H. M. 1984, \apj, 285, 89

\bibitem[]{}
Eisenhardt, P. R., Armus, L., Hogg, D. W., Soifer, B. T., Neugebauer,
G., Werner, M. W. 1996, \apj, 461, 72 

\bibitem[]{}
Elston, R., McCarthy, P. J., Eisenhardt, P.,
Dickinson, M., Spinrad, H., Januzzi, B. T., \& Mahoney, P. 1994,
\aj, 107, 910

\bibitem[]{}
Evans, A. S., Sanders, D. B., Mazzarella, J. M., Solomon, P. M.,
Downes, D., Kramer, C., \& Radford, S. J. E. 1996, \apj, 457, 658

\bibitem[]{}
Graham, J. R. \& Liu, M. C. 1995, \apjl, 449, L29 

\bibitem[]{}
Guelin, M., Kramer, C., \& Wild, W. 1995, IRAM Newsletter, 19, 17

\bibitem[]{}
Guilloteau, S. et al. 1997, A\&A, in press

\bibitem[]{}
G\"{u}sten, R., Serabyn, E., Kasemann, C., Schinckel, A., Schneider,
G.,
Schulz, A., \& Young, K. 1993, \apj, 402, 537

\bibitem[]{}
Hill, G. J., Thompson, K. L., \& Elston, R. 1993, \apjl, 414, 1

\bibitem[]{}
Hines, D. C. 1991, \apjl, 374, L9

\bibitem[]{}
Hines, D. C. 1994, Ph.D. Thesis, University of Texas

\bibitem[]{}
Hines, D. C., Schmidt, G. D., Smith, P. S., Cutri, R. M.,
\& Low F. J. 1995, \apjl, 450, L1

\bibitem[]{}
Hines, D. C. \& Wills, B. J. 1993, \apj, 415, 82

\bibitem[]{}
Hodapp, K., Hora, J. L., Irwin, E., \& Young, T. 1994, \pasp, 106, 87

\bibitem[]{}
Hodapp, K., Hora, J. L., Hall, D. N., Cowie, L. L., et al. 1996,
New Astronomy, 1, 176

\bibitem[]{}
Jannuzi, B. T., Elston, R., Schmidt, G. D., Smith, P. S.,
\& Stockman, H. S. 1994, \apj, 429, L49

\bibitem[]{}
Jim, K. T. C. et al. 1997, \pasp, in preparation

\bibitem[]{}
Kay, L. E. \& Miller, J. S. 1989, \baas, 21, 1099
 
\bibitem[]{}
Kim, D.-C. 1995, Ph.D. Thesis, University of Hawaii

\bibitem[]{}
Kim, D.-C., Sanders, D. B., Veilleux, S., Mazzarella, J. M.,
\& Soifer, B. T. 1995, \apjs, 98, 129 

\bibitem[]{}
Kleinmann, S. G., Hamilton, D., Keel, W. C., Wynn-Williams, C. G.,
Eales, S. A., Becklin, E. E., \& Kuntz, K. D. 1988, \apj, 328, 161

\bibitem[]{}
Kleinmann, S. G., \& Keel, W. C. 1987,
in Star Formation in Galaxies, ed. Carol J. Lonsdale Persson
(NASA C.P. 2466), 559

\bibitem[]{}
Liu, M. C., Graham, J. R., \& Wright, G. S. 1996, ApJ, 470, 771

\bibitem[]{}
Matthews, H. E. 1992, in The James Clerk Maxwell Telescope:
A Guide for the Prospective User,
ed. H. E. Matthews (Joint Astronomy Center, Hilo)

\bibitem[]{}
Moshir, M., et al. 1992, Explanatory Supplement 
to the IRAS Faint Source Catalog, 
Version 2, JPL D-10015 8/92 (Pasadena: JPL) (FSC)

\bibitem[]{}
Ohta, K., Yamada, T., Nakanishi, K, Kohno, K., Akiyama, M.,
\& Kawabe, R. 1996, Nature, 382, 426

\bibitem[]{}
Omont, A., Petitjean, P., Guilloteau, S., McMahon, R. G., Solomon,
P. M., \& Pecontal, E. 1996a, Nature, 382, 428

\bibitem[]{}
Omont, A., McMahon, R. G., Cox, P., Kreysa, E., Bergeron, J.,
Pajot, F., \& Storrie-Lombardi, L. J. 1996b, A\&A, 315, 1 

\bibitem[]{}
Osterbrock, D. E., Tran, H. D., Veilleux, S. 1992, \apj, 389, 196

\bibitem[]{}
Pozzetti, L., Madau, P., Zamorani, G., Ferguson, H. C., \& Bruzual,
G. A. 1998, MNRAS, in press

\bibitem[]{}
Sanders, D. B. 1991, in Dynamics of Galaxies and Their Molecular Cloud
Distributions, eds F. Combes and F. Casoli (Dordrecht: Kluwer), 417

\bibitem[]{}
Sanders, D. B. \& Mirabel, I. F. 1985, \apjl, 298, L31

\bibitem[]{}
Sanders, D. B. \& Mirabel, I. F. 1996, \araa, 34, 749

\bibitem[]{}
Sanders, D. B., Phinney, E. S., Neugebauer, G., Soifer, B. T., \&
Matthews, K. 1989, \apj, 347, 29

\bibitem[]{}
Sanders, D. B., Scoville, N. Z., \& Soifer, B. T. 1988c, \apjl, 335, L1

\bibitem[]{}
Sanders, D. B., Scoville, N. Z., \& Soifer, B. T. 1991, \apj, 370, 158

\bibitem[]{}
Sanders, D. B., Scoville, N. Z.,
Tilanus, R. P. J., Wang, Z., \& Zhou, S.
1993,
in Back to the Galaxy, eds S. Holt and F. Verter (New York: AIP), 311

\bibitem[]{}
Sanders, D. B., Soifer, B. T., Elias, J. H., Madore, B. F., 
Matthews, K., Neugebauer, G., \& Scoville, N. Z. 1988a,
\apj, 325, 74

\bibitem[]{}
Sanders, D. B., Soifer, B. T.,
Elias, J. H., Neugebauer, G., \& Matthews,
K.
1988b, \apjl, 328, L35

\bibitem[]{}
Sanders, D. B., Young, J. S., Scoville, N. Z., Soifer, B. T.,
\& Danielson, G. E. 1987, \apjl, 312, L5

\bibitem[]{}
Schmidt, M. \& Green, R. F. 1983, \apj, 269, 352

\bibitem[]{}
Scoville, N. Z., Padin, S., Sanders, D. B., Soifer, B. T., 
\& Yun, M. S. 1993, \apjl, 415, L75

\bibitem[]{}
Scoville, N. Z., \& Sanders, D. B. 1987,
in Interstellar Processes, eds
D. Hollenbach and H. Thronson (Dordrecht: Reidel), 21

\bibitem[]{}
Scoville, N. Z., Sanders, D. B., Sargent, A. I., Soifer, B. T.,
Scott, S. L., \& Lo, K. Y. 1986, \apjl, 311, L47

\bibitem[]{}
Scoville, N. Z., Yun, M. S., Windhorst, R. A., Keel, W. C., 
\& Armus, L. 1997, ApJ, 485, L21

\bibitem[]{}
Serjeant, S., Lacy, M., Rawlings, S., King, L. J.,
\& Clements, D. L. 1995, \mnras, 276, L31 

\bibitem[]{}
Soifer, B. T., Cohen, J. G., Armus, L., Matthews, K.,
Neugebauer, G., \& Oke, J. B. 1995, \apj, 443, L65

\bibitem[]{}
Soifer, B. T., Neugebauer, G., Armus, L., Shupe, D. L. 1996, \aj, 111,649

\bibitem[]{}
Soifer, B. T., Neugebauer, G., Matthews, K., \& Armus, L. 1994, \apjl,
433, L69

\bibitem[]{}
Solomon, P. M., Downes, D., \& Radford, S. J. E. 1992a, \apj, 398, L29

\bibitem[]{}
Solomon, P. M., Radford, S. J. E., \& Downes, D. 1992b,
Nature, 356, 318

\bibitem{}{}
Surace, J.A., Sanders, D.B. Vacca, W.D., Veilleux, S.,
\& Mazzarella, J.M. 1998, \apj, 492, 116

\bibitem[]{}
Trentham, N. 1995, \mnras, 277, 616 

\bibitem[]{}
Tsuboi, M. \& Nakai, N. 1992, \pasj, 44, L241

\bibitem[]{}
Tytler, D. \& Fan, X. 1992, \apjs, 79, 1

\bibitem[]{}
Veilleux, S., Kim, D.-C., Sanders, D. B.,
Mazzarella, J. M., \& Soifer, B. T. 1995, \apjs, 98, 171 

\bibitem[]{}
Veilleux, S. \& Osterbrock, D. E. 1987, \apjs, 63, 295

\bibitem[]{}
Wainscoat, R. J. 1996, University of Hawaii Telescopes at Telescopes at
Mauna Kea Observatory - User Manual, University of Hawaii

\bibitem[]{}
Wink, J. E., Guilloteau, S., \& Wilson, T. L.
1997, A\&A, 322, 427

\bibitem[]{}
Young, J. S., Kenney, J., Lord, S., \& Schoerb, F. P. 1984,
\apjl, 287, L65

\bibitem[]{}
Yun, M. S., Scoville, N. Z., \& Evans, A. S. 1998, in
Highly Redshifted Radio Lines, eds. C. Carilli, S. J. E.
Radford, K. Menten, \& G. Langston, in press

\end{thebibliography}
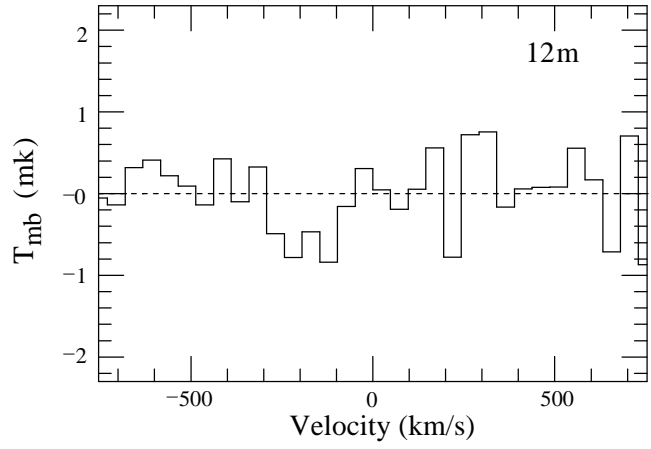
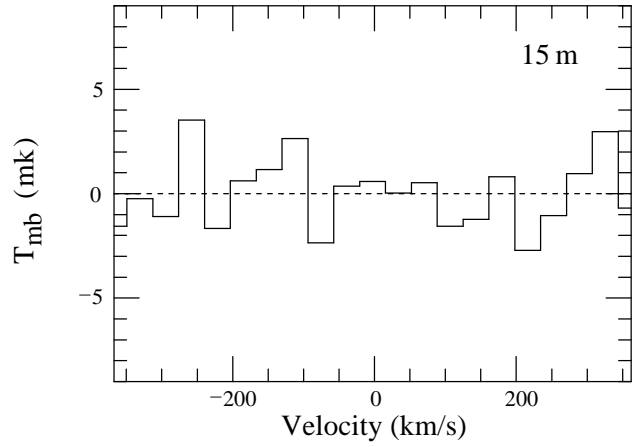
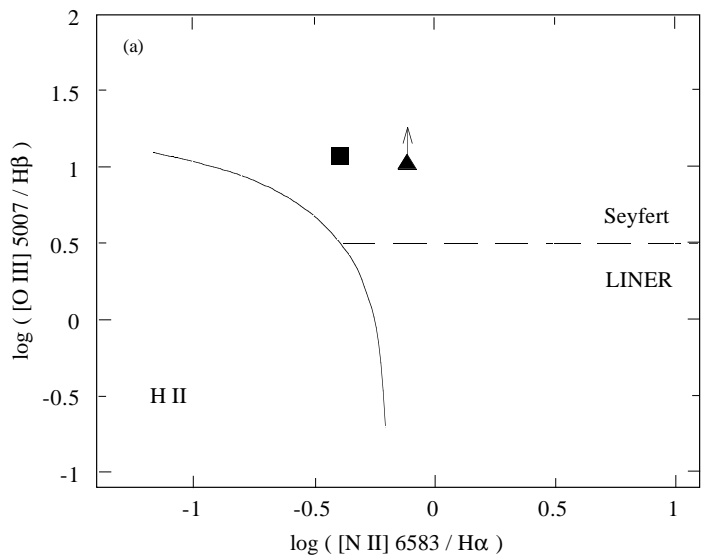
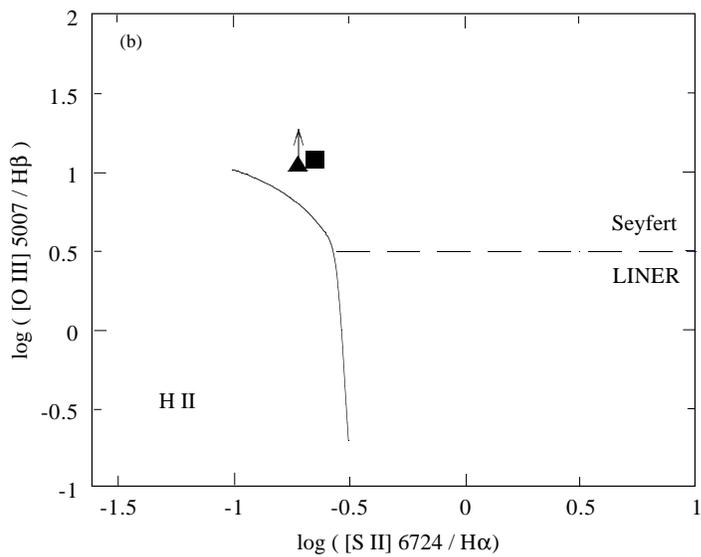
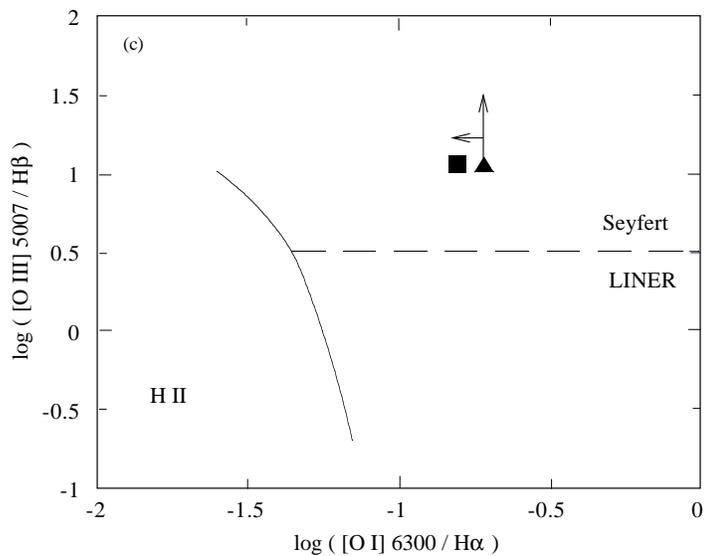
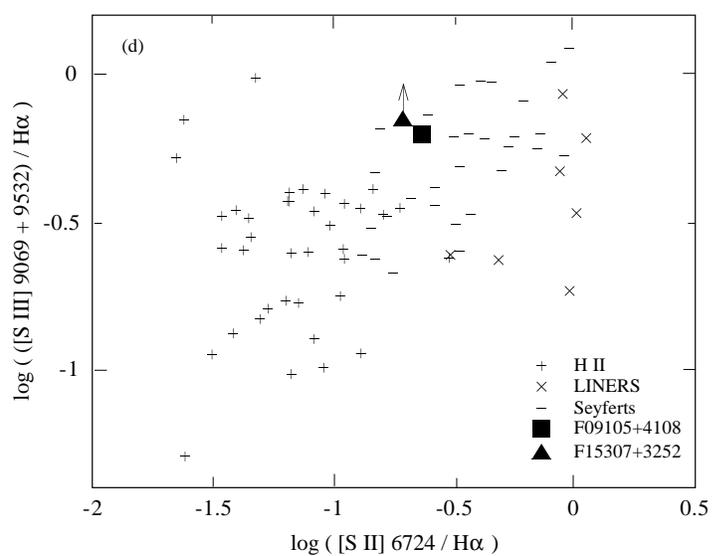
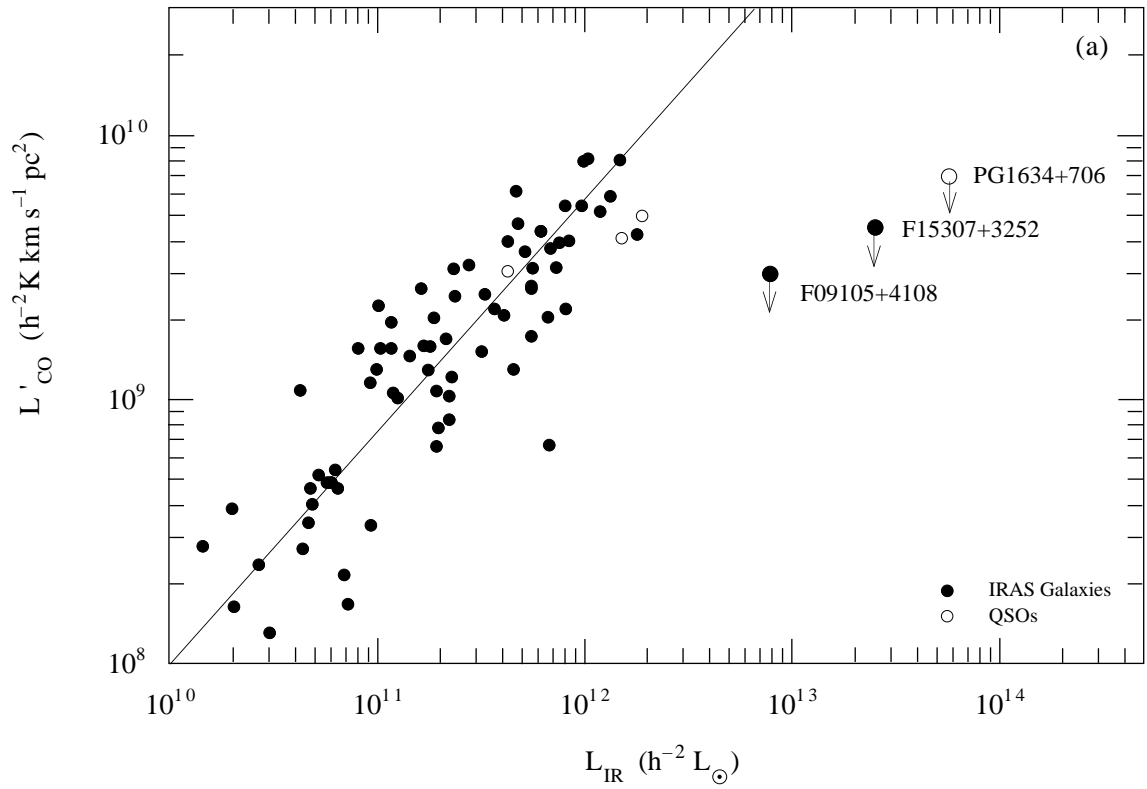
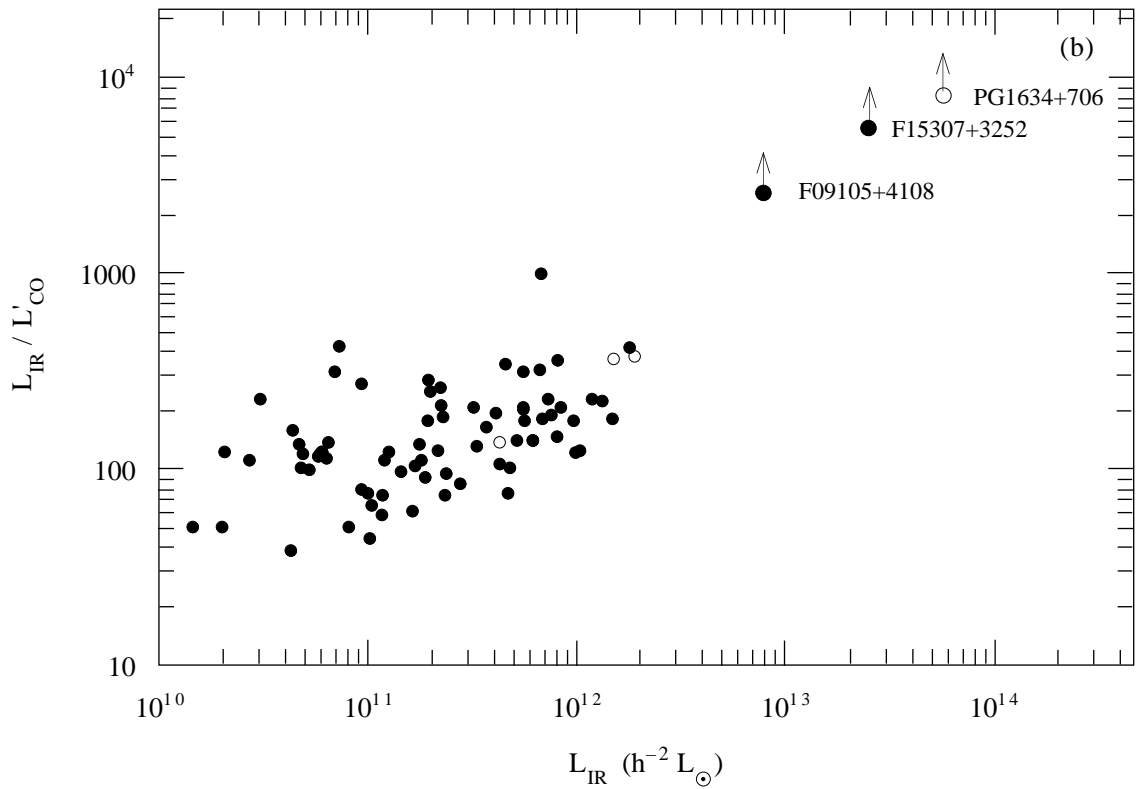
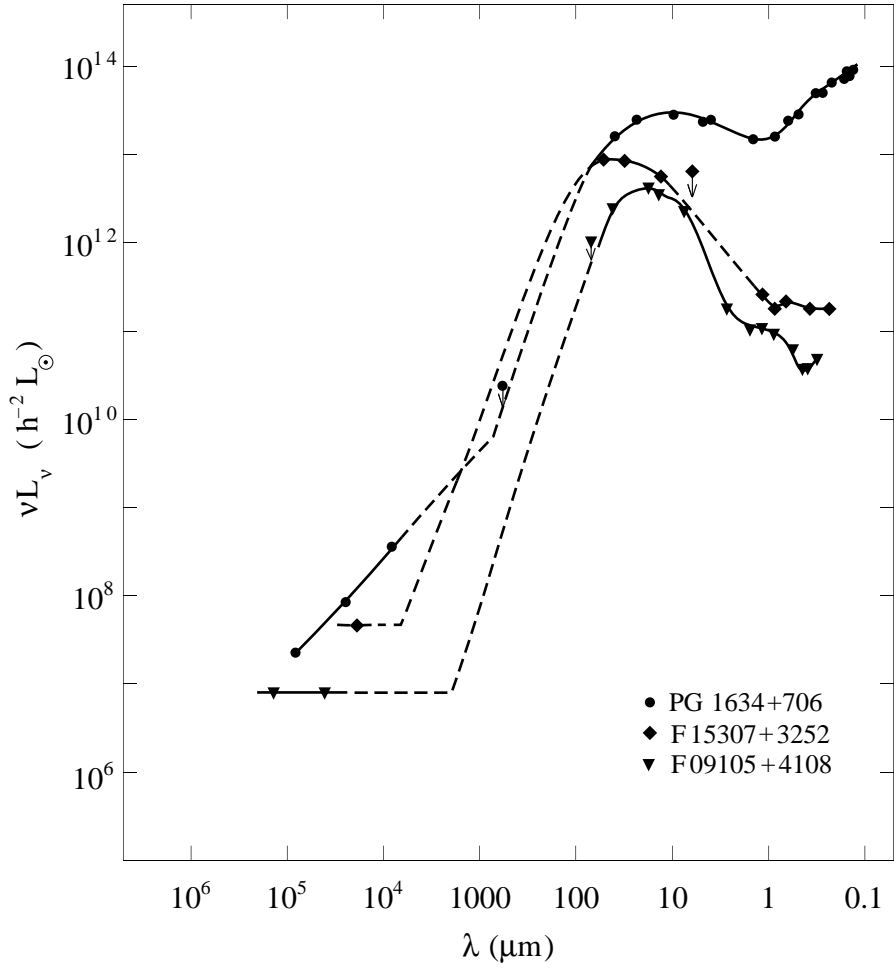
\end{document}